Current-Induced Magnetization Switching in MgO Barrier Based Magnetic Tunnel Junctions with CoFeB/Ru/CoFeB Synthetic Ferrimagnetic Free Layer

Jun HAYAKAWA[1,2], Shoji IKEDA[2], Young Min LEE[2], Ryutaro SASAKI[2],

Toshiyasu MEGURO[2], Fumihiro MATSUKURA[2], Hiromasa TAKAHASHI[1,2]

and Hideo OHNO[2]

[1]*Hitachi, Ltd., Advanced Research Laboratory, 1-280 Higashi-koigakubo, Kokubunji, Tokyo 185-8601, Japan*

[2]*Laboratory for Nanoelectronics and Spintronics, Research Institute of Electrical Communication, Tohoku University, 2-1-1 Katahira, Aoba-ku, Sendai 980-8577, Japan*

We report the intrinsic critical current density ($J_{c0}$) in current-induced magnetization switching and the thermal stability factor ($E/k_\mathrm{B}T$, where $E$, $k_\mathrm{B}$, and $T$ are the energy potential, the Boltzmann constant, and temperature, respectively) in MgO based magnetic tunnel junctions with a $Co_{40}Fe_{40}B_{20}$(2 nm)/Ru(0.7-2.4 nm)/$Co_{40}Fe_{40}B_{20}$(2 nm) synthetic ferrimagnetic (SyF) free layer. We show that $J_{c0}$ and $E/k_\mathrm{B}T$ can be determined by analyzing the average critical current density as a function of coercivity using the Slonczewski's model taking into account thermal fluctuation. We find that high antiferromagnetic coupling between the two CoFeB layers in a SyF free layer results in reduced $J_{c0}$ without reducing high $E/k_B T$.

Spin-polarized currents exert torque on a magnetization that can switch the magnetization direction once the current density becomes sufficiently high.[1,2)] Such a current-induced magnetization switching (CIMS) at reduced current density has been demonstrated in a number of magnetic tunnel junctions (MTJs).[3-5)] In particular, CoFeB/MgO/CoFeB MTJs have been shown to exhibit high tunnel magnetoresistance (TMR) ratios together with CIMS.[6-9)] In order to use this system for magnetic random access memories, however, further reduction of the critical current density ($J_c$) while maintaining a high thermal stability factor over 40 ($E/k_BT$, where $E$, $k_B$, and $T$ are the energy potential, the Boltzmann constant, and temperature, respectively) is required to satisfy the nonvolatility over 10 years. Meeting this requirement with a conventional single ferromagnetic free layer patterned into a nanometer-scale shape appears to be not easy as the thermal stability degrades as the dimension is reduced. Synthetic ferrimagnetic (SyF) free layer has been investigated as an alternative to the single free layer because the SyF structure consisting of two or more ferromagnetic layers separated by a non-magnetic spacer such as Ru film is expected to provide high volume to withstand thermal fluctuations.[10-12)] In current-perpendicular-to-plane giant magnetoresistance (CPP-GMR) nanopillars, where only a small magnetoresistance ratio can be obtained, CIMS with SyF free layers were investigated and $J_c$, measured using a dc sweep, was reported to be lower than that of conventional single free layer MTJs[13].

However, intrinsic critical current density $J_{c0}$, which we define in the following, and $E/k_BT$ of such a nanopillar, either CPP-GMR or MTJ, with a SyF free layer have not been investigated so far.

In this letter, we focus on MgO barrier based MTJs with a $Co_{40}Fe_{40}B_{20}/Ru/Co_{40}Fe_{40}B_{20}$ SyF free layer and investigate $J_{c0}$ and $E/k_BT$. The $J_{c0}$ and $E/k_BT$ are shown to be able to be determined by analyzing the average critical current density ($J_c^{ave}$) as a function of coercivity using the Slonczewski's model taking into account thermal fluctuation. Thus obtained $J_{c0}$ and $E/k_BT$ are in good agreement with those obtained by $J_c^{ave}$ vs $\ln(\tau_p/\tau_0)$ plots, where $\tau_p$ and $\tau_0$ represent the current pulse duration and inverse of the precession frequency. The $J_{c0}$ is defined as the critical current density at the inverse precession frequency.

Figure 1(a) is a schematic diagram of the MTJ pillar we fabricated. MTJ films were deposited on $SiO_2$/Si substrates by using RF magnetron sputtering with a base pressure of $10^{-9}$ Torr. The order of the film layers was, starting from the substrate side, Ta(5)/Ru(50)/Ta(5)/NiFe(5)/MnIr(8)/CoFe(4)/Ru(0.8)/$Co_{40}Fe_{40}B_{20}$(5)/MgO(0.9-1.0)/$Co_{40}Fe_{40}B_{20}$(2)/Ru($t_{Ru}$)/$Co_{40}Fe_{40}B_{20}$(2)/Ta(5)/Ru(5) (in nm). The Ru spacers varying in thickness from 0.7 to 2.4 nm were formed by using a slide mask shutter during sputtering. All nano-scaled junctions were fabricated using an electron-beam lithography process. Figure 1(b) shows scanning electron microscopy images of rectangular-shaped MTJ pillars with the dimension of 80 x 160 nm$^2$ (top image) and 80 x 240 nm$^2$ (bottom one). The completed MTJs were annealed at temperature of 300 $^o$C for 1 h in a 10$^{-6}$ Torr vacuum under a magnetic field of 4 kOe. The TMR loops of the MTJs were measured at room temperature using a four-probe method with dc bias and magnetic field of up to 1 kOe. CIMS was evaluated by measuring resistance by 50 μA-step current pulses with the pulse duration ($\tau_p$) ranging from 100 μs to 1 s. The thermal stability factor $E/k_BT$ was obtained from the slope of the average critical current density $J_c^{ave}$ vs ln($\tau_p/\tau_0$) plot, in addition to the method proposed in this paper. The current direction is defined as positive when the electrons flow from the top (free) to the bottom (pin) layer.

In order to determine the magnetic exchange coupling energy $J_{ex}$ between the two CoFeB ferromagnetic layers as a function of $t_{Ru}$ in the CoFeB(2)/Ru($t_{Ru}$)/CoFeB(2) structure, we prepared separately a structure, SiO$_2$/Si substrate /Ta(5)/Ru(50)/Ta(5)/MgO(0.9)/CoFeB(2)/Ru($t_{Ru}$)/CoFeB(2)/Ta(5), which is cut into a rectangular with the dimension of 1 x 3 mm$^2$ for magneto-optical Kerr effect measurements. Figure 2(a) plots magnetic exchange coupling energy $J_{ex}$ between the two CoFeB ferromagnetic layers as a function of $t_{Ru}$. Inset shows the expanded view of the range from $t_{Ru}$ = 1.0 to 3.0 nm. We have calculated $J_{ex}$ using $J_{ex}$ = $-\mu_0 H_s M_s t_1 M_2 t_2/(M_1 t_1 + M_2 t_2)$, where $H_s$ is saturation field, $M_{1,2}$ saturation magnetization of CoFeB (1.3 T), $t_{1,2}$ thickness of CoFeB (2 nm).[4,14] The highest antiferromagnetic coupling energy of 0.17 mJ/m$^2$ is obtained at $t_{Ru}$ ~ 0.6 nm. We also see oscillations in the magnitude of $J_{ex}$; the second peak is located at $t_{Ru}$ = 1.2 nm and the third at $t_{Ru}$ = 2.4 nm. The oscillatory behavior of $J_{ex}$ in Fig. 2(a) suggests the presence of an oscillation from ferromagnetic to antiferromagnetic and back as reported earlier[15], which originates from the Ruderman-Kittel-Kasuya-Yosida (RKKY)-type coupling typically found in Co/Ru/Co multilayers.[16] Note that the open circles in Fig. 2(a) represent either no $J_{ex}$ or positive (ferromagnetic) $J_{ex}$, because we cannot measure the ferromagnetic coupling by the method employed here. Figure 2(b) shows the coercivity $H_c$ obtained from the TMR measurements under magnetic field as a function of $t_{Ru}$ of the nanoscaled MTJs. The $H_c$ strongly depends on the $t_{Ru}$ and greater the $J_{ex}$ higher the $H_c$.

Figures 3(a) and 3(b) show the magnetic field hysteresis loop (*R-H* loop) and the resistance versus pulsed current (*R-I*$_p$) with $\tau_p$ = 10 ms of a 80 x 160 nm$^2$ MTJ with a CoFeB(2)/Ru(0.7)/CoFeB(2) SyF free layer. The Ru thickness of 0.7 nm corresponds to the highest antiferromagnetic coupling energy. The TMR ratio is 90%, which is comparable to the one reported for an MTJ with a 2-nm CoFeB single free layer.[4)] The $R$-$I_p$ curves were measured under an applied magnetic field of -32 Oe along the direction of the pin CoFeB layer to compensate the offset field [see Fig. 3(a)] arising primarily from the stray fields of the edge of the patterned SyF pin layer. The current density required to switch the magnetization from parallel (anti-parallel) to anti-parallel (parallel) shown in Fig. 3(b) is $J_c^{P\rightarrow AP}$ = 6.8 x 10$^6$ A/cm$^2$ ($J_c^{AP\rightarrow P}$ = -6.8 x 10$^6$ A/cm$^2$); the average critical current density ($J_c^{ave.}$), defined as ($|J_c^{P\rightarrow AP}|+|J_c^{AP\rightarrow P}|$)/2, is 6.8 x 10$^6$ A/cm$^2$. Figure 3(c) plots $J_c^{ave}$ as a function of ln($\tau_p/\tau_0$) for $\tau_p$ from 100 μs to 1 s for the same MTJs shown in Figs. 3(a) and 3(b). Based on eq. (1) shown later, the slope of this plot reveals the thermal stability factors $E/k_BT$ to be 67. By extrapolating $J_c$ to ln($\tau_p/\tau_0$) of 0 which corresponds to $\tau_p$ = 1 ns, we obtain the intrinsic critical current density ($J_{c0}$) of 8.7 x 10$^6$ A/cm$^2$.

Figure 4(a) plots $J_c^{ave}$ as a function of 1/$H_c$ (inverse of the coercivity $H_c$) of all the MTJs investigated in this study, with varying Ru spacers from 0.7 to 2.4 nm. Here, all $J_c^{ave}$ were measured with a pulse current of 1 s duration. We have noticed that (1) $J_c^{ave}$ can be categorized into three groups, depending on the strength of the magnetic coupling energy $J_{ex}$ shown in Fig. 2, and (2) within each group $J_c^{ave}$ increases linearly with $H_c$. The black symbols ($t_{Ru}$ = 0.7 and 0.9 nm), white symbols ($t_{Ru}$ = 1.5, 1.7, and 1.9 nm), and hatched square symbols ($t_{Ru}$ = 2.2 and 2.4 nm) correspond to the Ru spacer thickness for the first antiferromagnetic coupling, the ferromagnetic coupling between the second and third anti-ferromagnetic coupling, and the third antiferromagnetic coupling, respectively. Hereafter, we call them Group I, Group II, and Group III, respectively.

We now show that the $J_{c0}$ and $E/k_BT$ can be obtained by analyzing the measured $J_c^{ave}$ versus $H_c$ using the following Slonczewski's model[1,17] taking into account the thermal activated nature of the magnetization switching. The relevant equations are;[18,19]

$$J_c = J_{c0}\{1-(k_BT/E)\ln(\tau_p/\tau_0)\}, \quad (1)$$

$$J_{c0} = \alpha\gamma eM_st(H_{ext} \pm H_k \pm H_d)/\mu_B g, \quad (2)$$

$$E = M_sVH_k/2, \quad (3)$$

$$g = P/[2(1+P^2\cos\theta)], \quad (4)$$

where $\alpha$ is the Gilbert damping coefficient, $\gamma$ the gyromagnetic constant, $e$ the elementary charge, $t$ the thickness of the free layer, $H_{ext}$ the external magnetic field, $H_k$ the in-plane uniaxial magnetic anisotropy, $M_s$ the saturation magnetization of free layer, $V$ the volume of free layer and $H_d$ the out-of-plane magnetic anisotropy induced by the demagnetization field. $\theta$ is 0 for the parallel configuration and $\pi$ for anti-parallel. In the following, we assume that our magnetic cell has a uni-axial anisotropy and a single magnetic domain, and hence $H_k \approx H_c$, because we are dealing with nanoscale structures. Since $H_d \gg H_k, H_{ext}$, we obtain the average $J_{c0}$ ($J_{c0}^{ave.}$), and $J_c^{ave}$, which is a function of $H_c$, as,

$$J_{c0}^{ave} \approx \alpha \gamma e M_s t H_d (g^{(P \to AP)} + g^{(AP \to P)}) / \mu_B g^{(P \to AP)} g^{(AP \to P)} \tag{5}$$

$$J_c^{ave} = J_{c0}^{ave} [1-(2k_B T/M_s V) \ln(\tau_p/\tau_0)/H_c] \tag{6}$$

As seen in Fig. 4(a), the measured $J_c^{ave}$ are inversely proportional to $H_c$ within each group (solid lines). This shows that eq. (6) is a good approximation describing our results. By extrapolating $1/H_c$ to zero in Fig. 4(a), we can then obtain $J_{c0}^{ave}$, given by eq. (5), as 9.4 x 10$^6$ A/cm$^2$ for Group I, 1.57 x 10$^7$ A/cm$^2$ for Group II, and 1.65 x 10$^7$ A/cm$^2$ for Group III. Thus determined $J_{c0}^{ave}$ value for Group I is in good agreement with the value obtained from the $J_c^{ave}$ versus $\ln(\tau_p/\tau_0)$ plot of Fig. 3(c) (8.7 x 10$^6$ A/cm$^2$). This also shows that $\tau_0 = 1$ ns is a reasonable value for the inverse precession frequency. We speculate that the reduction of $J_{c0}^{ave}$ for the MTJs of Group I may be related to spin-accumulation; two antiferromagnetically coupled CoFeB layers separated by a non-magnetic Ru layer whose thickness is much thinner than the spin diffusion length[20] is known to enhance the spin accumulation at the CoFeB and Ru interface.[21,22] Spin accumulation can increase the efficiency of spin-torque acting on the CoFeB free layer and contribute to the reduction of critical current density. Clearly more work is needed to clarify the mechanism.

Next, we compare the thermal stability factors $E/k_BT$ between MTJs categorized into three groups from Fig. 4(a). The slope of $J_c^{ave}$ vs. $1/H_c$ is given by $-J_{c0}^{ave}(2k_BT/M_sV)\ln(\tau_p/\tau_0)$ according to eq. (6). Because $E/k_BT = M_sVH_c/2k_BT$ the factor obtained by dividing the slope by $-J_{c0}^{ave}(\ln(\tau_p/\tau_0))$ x $H_c$ should yield the $E/k_BT$, where $\tau_p$ and $\tau_0$ are 1 s and 1 ns, respectively. We calculate the factors $(2k_BT/M_sV)^{-1}$ x $H_c$ for all the MTJs shown in Fig. 4(a) as summarized in Fig. 4(b), where we employ the $J_{c0}^{ave}$ obtained by extrapolating $1/H_c$ to zero in Fig. 4(a). We notice that the MTJs having a large antiferromagnetic coupling energy (Group I) exhibit smaller $J_{c0}$ values together with higher $(2k_BT/M_sV)^{-1}$ x $H_c$, hence higher $E/k_BT$, than those in other groups (Group II and Group III).

In conclusion, we have determined intrinsic critical current density ($J_{c0}$) for current-induced magnetization switching and the thermal stability factor ($E/k_BT$) of MgO barrier based MTJs with $Co_{40}Fe_{40}B_{20}$(2 nm)/Ru(0.7-2.4 nm)/$Co_{40}Fe_{40}B_{20}$(2 nm) synthetic ferromagnetic (SyF) free layers. The $J_{c0}$ and $E/k_BT$ are shown to be able to be determined using $J_c^{ave}$ vs $1/H_c$ curves. High antiferromagnetic coupling between the two CoFeB ferromagnetic layers is found to result in a reduced $J_{c0}$ together with a high thermal stability factor $E/k_BT$ over 60.

This work was supported by the IT-program of the Research Revolution 2002 (RR2002): "Development of Universal Low-power Spin Memory", Ministry of Education, Culture, Sports, Science and Technology of Japan.

Mater. **3** (2004) 361.

Figure captions

Fig. 1. (a) Schematic drawing of the cross section of the MTJs. The thickness of the Ru spacer in the SyF layer varies from 0.7 to 2.4 nm. (b) Scanning electron microscopy image of the pillars.

Fig. 2. (a) Magnetic exchange coupling energy $J_{ex}$ for CoFeB/Ru/CoFeB SyF layers with Ru spacers varying from 0 to 3 nm in thickness. Inset shows the expanded view of the plots ranging from $t_{Ru}$ = 1.0 to 3.0 nm. (b) Coercivity ($H_c$) as a function of $t_{Ru}$ in the nano-scaled MTJs with CoFeB/Ru/CoFeB SyF free layer.

Fig. 3. $R$-$H$ loops (a), $R$-$I_p$ loops at $\tau_p$ of 10 ms (b), and $J_c^{ave}$ [($J_c^{P\to AP}$-$J_c^{AP\to P}$)/2] as functions of $\ln(\tau_p/\tau_0)$ (c) at room temperature for an MTJ with a CoFeB(2 nm)/Ru(0.7 nm)/CoFeB(2 nm) SyF free layer.

Fig. 4. $J_c^{ave}$ as a function of $1/H_c$ (inverse of the $H_c$) (a), and $E/k_B T$ (b) for MTJs with SyF free layers having Ru spacers ranging from 0.7 to 2.4 nm. The plotted black symbols ($t_{Ru}$ = 0.7, 0.9, and 1.2 nm), white symbols ($t_{Ru}$ = 2.2 and 2.4 nm), and hatched

square symbols ($t_{Ru}$ = 1.5, 1.7, and 1.9 nm) correspond to the Ru spacer thickness for the first/second antiferromagnetic coupling, the ferromagnetic coupling between the second and third antiferromagnetic coupling, and the third antiferromagnetic coupling, respectively.

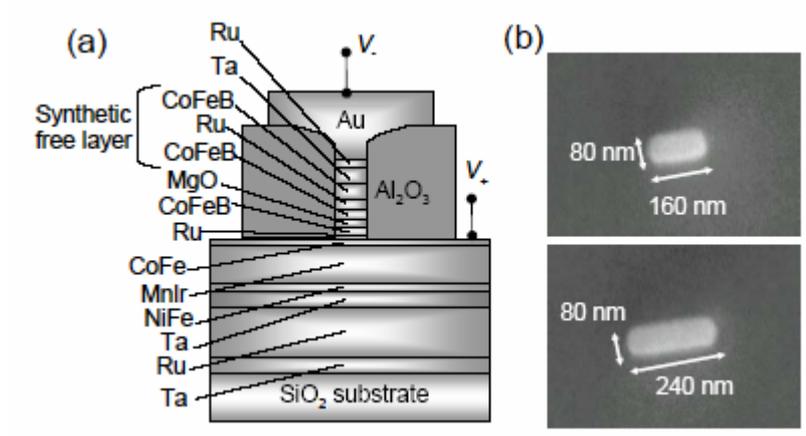

Fig. 1  Hayakawa  et al.

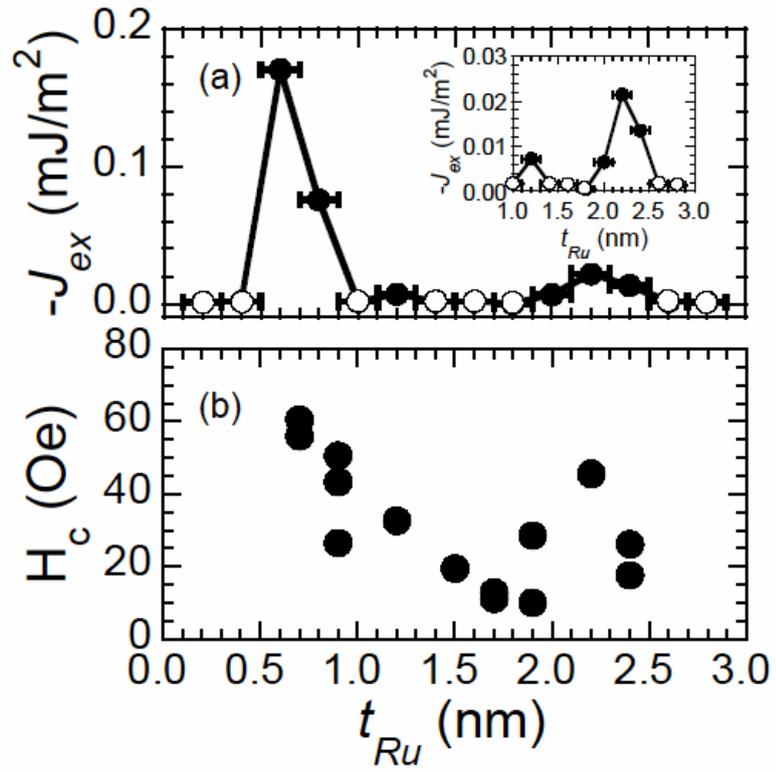

Fig. 2   Hayakawa   et al.

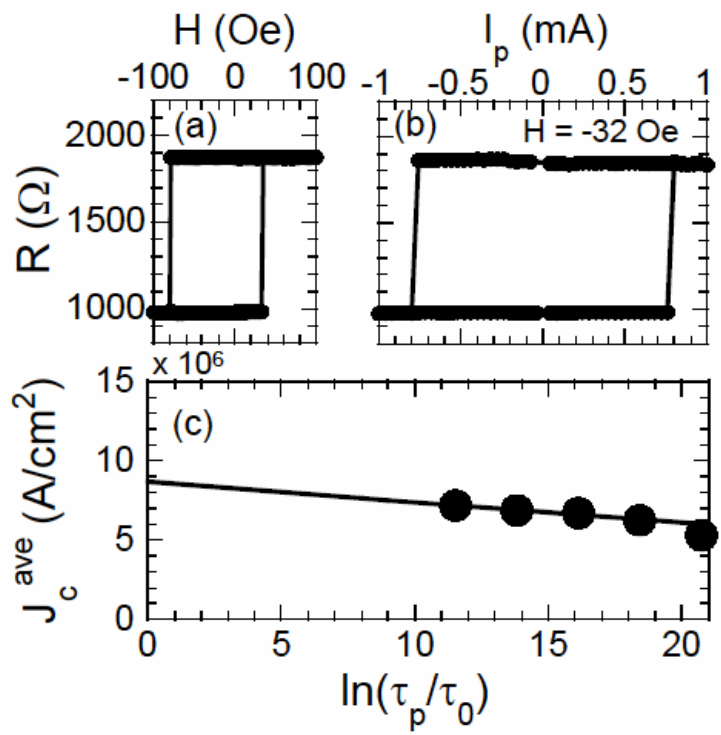

Fig. 3   Hayakawa   et al.

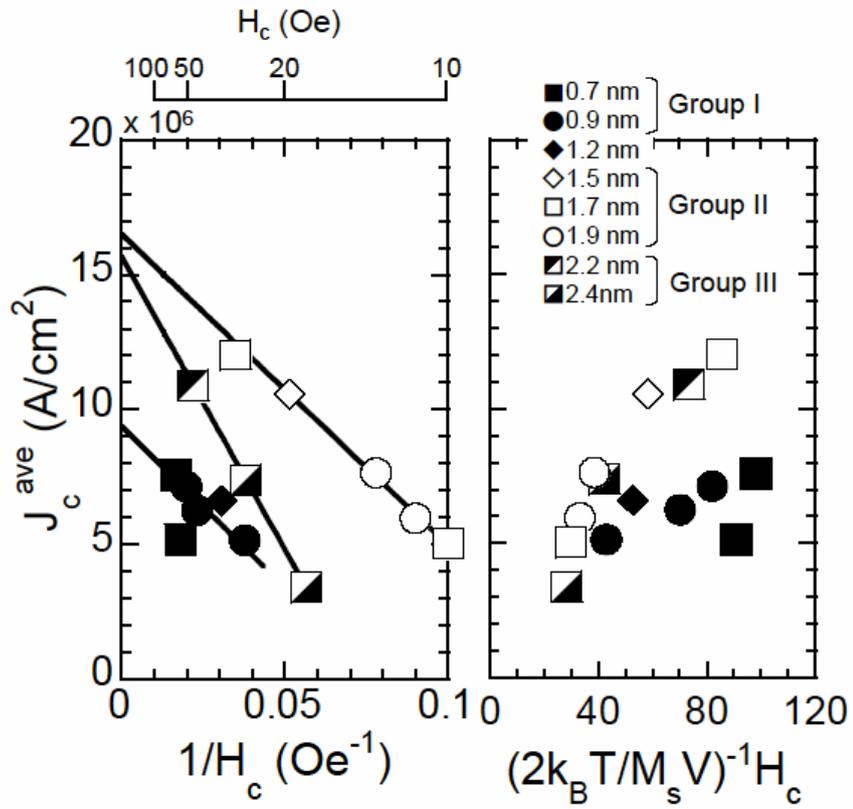

Fig. 4   Hayakawa   et al.